      [1999/12/01 v1.4c Il Nuovo Cimento]
\documentclass{cimento}

\newcommand{\be}{\begin{eqnarray}}
\newcommand{\ee}{\end{eqnarray}}

\title{Contemporary gravitational waves from primordial black holes
}
\author{A.D.~Dolgov\from{ins:x}
}
\instlist{\inst{ins:x} University of Ferrara and INFN, via Saragat, 1, 44100, Italy; \\
 ITEP, Moscow, 117218, Russia                         }
\PACSes{\PACSit{00.00}
\PACSit{04.30.-w, 98.80.-k, 14.65.Jk	}
}
\begin{document}

\maketitle

\begin{abstract}
Stochastic background of gravitational waves (GW) generated by the interactions between 
primordial black holes (PBH) in the early universe and by PBH evaporation is considered. 
If PBHs dominated in the cosmological energy density prior to their evaporation, GWs from 
the earlier stages (e.g. inflation) would be noticeably diluted. On the other hand, at the PBH 
dominance period they could form dense clusters where PBH binary formation might be 
significant. These binaries would be efficient sources of the gravitational waves.
\end{abstract}

The registration of the gravitational waves generated in the early universe could 
bring an important information about inflation, possible (first order) cosmological phase
transitions, topological defects, such as cosmic strings, etc. Very sensitive GW detectors
such as LIGO and LISA may make make the discovery opening a new era of gravitational
wave astronomy. In particular, an observation of stochastic cosmological background of low 
frequency GWs could be a final proof of inflation. However, an absence of a such background
would not mean that the universe was not in inflationary stage. First, the present day density 
of GWs depends upon the model of inflation and, second, there could be a mechanism
which suppresses the density of GWs at post-inflationary stage. Such a mechanism is
described in my talk. Though the inflationary background of GWs could be noticeably 
suppressed, a new higher frequency GWs would be generated by the suggested mechanism.
The talk is based on two papers~\cite{ref:dnn,ref:de}. In the second one a detailed reference list is
presented which is reduced here due to lack of space.

We consider GWs produced by the interactions between primordial black holes (PBH), as well as 
by their evaporation. PBHs are supposed to be very light, so they evaporated before the big bang 
nucleosynthesis (BBN) leaving no trace in the present day universe, except for GWs. The life-time
of an evaporating black hole with mass $M$ is equal to~\cite{ref:pbh-life}
\be 
\tau_{BH} = \frac{10240\,\pi}{N_{eff}}\, \frac{M^3}{m_{Pl}^4}\,,
\label{tau-BH}
\ee
where $m_{Pl}=1.22\cdot 10^{19}\,{\rm GeV} = 2.176\cdot 10^{-5}$ g is the Planck mass
and $N_{eff}$ is the number of particle species with masses
smaller than the black hole temperature: 
\be
T_{BH}=\frac{m_{Pl}^2}{8\pi M}\, .
\label{T-BH}
\ee
The corrections due to the propagation and back-capture of the evaporated particles~\cite{ref:page},
the so called grey factor, change this result by a factor of order unity and are not included here.

According to ref.~\cite{ref:carr},
to avoid a conflict with BBN the life-time of PBHs should be shorter than
$t\approx 10^{-2}$ s and thus the black holes should be lighter than
\be
M < 1.75\cdot 10^8 \left(\frac{N_{eff}}{100}\right)^{1/3}\,\,{\rm g}.
\label{M-upper-bound}
\ee
The temperature of such PBHs exceeds $3\cdot 10^4 $ GeV and
correspondingly $N_{eff} \geq 10^2$.

Formation of PBHs from primordial density perturbations in the early
Universe was considered in pioneering papers
\cite{ref:zeld-nov,ref:hawk-bh}. PBHs  were formed 
when the density contrast, $\delta\rho /\rho$, at horizon was of the order unity or,
in other words, when the Schwarzschild radius of the perturbation was of the order of the
horizon scale. If PBHs were created at the radiation dominated stage, when the cosmological
energy density was  $\rho(t) = 3 m_{Pl}^2/(32\pi t^2) $, and the horizon was 
$l_h = 2t$, the mass of such PBHs would be:
\begin{equation}
M(t)={ m_{Pl}^2t}\simeq 4\cdot 10^{38} \,\left(\frac{t}{\rm{sec}}\right)\,{\rm g}\,,
\label{BH-mass}
\end{equation}
where $t$ is the cosmological time.

The fraction, $\Omega_p$, of the cosmological energy density of PBH produced 
by this mechanism depends upon the spectrum
of the primordial density perturbations. If the usual flat Harrison-Zeldovich spectrum is
assumed, then $\Omega_p$ would be quite small. We have not calculated $\Omega_p$ 
but have taken it as a free parameter of the model. One reason for that is that the spectrum of 
the density perturbations
at small wave lengths is unknown. Moreover, there could be other mechanisms of 
PBH formation. In particular, in 
refs.~\cite{ref:ad1992,ref:ad2008} a model of PBH formation has been proposed 
which might lead to considerably larger probability of PBH formation. 
The mass spectrum of PBHs produced by the latter mechanism
has the log-normal form:
\be
\frac{dN}{dM} = C \exp \left[\frac{(M-M_0)^2}{M_1^2}\right],
\label{dN-dM}
\ee 
where $C$, $M_0$, and $M_1$ are some model dependent parameters. Quite
naturally the central value of PBH mass distribution may be in the desired range $M_0 < 10^{9}$ g.
In this model the value of $\Omega_p$ may be much larger than in the conventional model
based on the flat spectrum of the primordial fluctuations. We will not further speculate on the
value of $\Omega_p$ and on the form of the mass spectrum of PBH. In what follows we assume
for an order of magnitude estimate that the spectrum is well localized near some fixed 
mass $M$ and that $\Omega_p$ is an arbitrary parameter. Different mechanisms of PBH production 
are reviewed e.g. in ref.~\cite{ref:carr-rev}.

The energy density of nonrelativistic PBHs drops down as $1/a^3$, while the 
energy density of the initially dominant relativistic matter drops as $1/a^4$, where
$a=a(t)$ is the cosmological scale factor. So the relative contribution of PBH into
the total energy density rises as 
\be
\Omega_{BH} (t) = \Omega_p {a(t)}/{a_p} = \Omega_p (t/t_p)^{1/2}\,,
\label{Omega-of-t}
\ee
here $t_p = M/m_{Pl}^2 $  is the PBH production time, related to their mass by eq. (\ref{BH-mass}).
So initially $\Omega_{BH} (t)$ rises as $t^{1/2}$ and at some stage it reaches unity and after that
$\Omega_{BH}$ remains constant till the PBH evaporation. PBHs would begin to dominate in the
cosmological energy density at $t =t_{eq} = {M}{(m_{Pl}^2 \Omega_p^2)}$, if 
$t_{eq} > \tau_{BH}$, eq.~(\ref{tau-BH}).  This can be transformed into the lower limit on the 
PBH mass:
\be
M > 6\cdot 10^{-2}\,\left(\frac{N_{eff}}{100}\right)^{1/2}\frac{m_{Pl} }{\Omega_p}
\simeq 10^{-7}\, {\rm g} \,\,\Omega_p^{-1}\,.
\label{M-for-t-MD}
\ee

If condition (\ref{M-for-t-MD}) was fulfilled, the universe expansion regime was initially
relativistic, radiation dominated (RD), then after $t=t_{eq}$ it became non-relativistic,
matter dominated (MD). Later after PBH evaporation, $t>\tau_{BH}$ the universe returned
to RD stage again, and only after very long time, $t =t_{LSS}\sim 10^5$ years, the expansion 
became matter dominated. After that time the large scale structures (galaxies, their clusters, etc)
began to form. As is known, cosmological structure formation took place at MD stage, when
initially small primordial density perturbations started to rise due to gravitational instability.
In our case the density perturbations  started to rise at $t>t_{eq}$. According to the theory,
at MD stage $\Delta \equiv \delta \rho /\rho \sim a(t)$, till the perturbations remains small,
$\Delta \ll 1$. When $\Delta$ reaches unity, the perturbations quickly rise and as a result
they becomes quite large, $\Delta \gg 1$. In the present day universe $\Delta \sim 10^5$ at the 
galactic scale. 

In our scenario we expect formation of high density clusters of PBHs with density contrast which 
rose as  $ \Delta (t) = \Delta _{in} (t/t_{in})^{2/3}$, where $t_{in}\geq t_{eq}$ is the moment when the
perturbation comes inside the cosmological horizon.
The density contrast would reach unity at  $t_1(t_{in})$ such that:
\be
\Delta [t_1(t_{in})]= \Delta_{in} [t_1(t_{in}) /t_{in}]^{2/3} =1\,\,\, 
\rm{or}\,\,\, t_1 (t_{in} )=t_{in} \Delta_{in}^{-3/2}\,.
\label{Delta-of-t1}
\ee 
To this end the PBH life-time should be longer than $t_1$.

After the density contrast has reached unity, the cluster would
decouple from the general cosmological expansion. In other words, 
the cluster stopped to expand
together with the universe and, on the opposite, it would begun to shrink 
when gravity took over the free streaming of PBHs. So the cluster size would drop down
and both $n_{BH}$ and $\rho_b$ would rise. The
density contrast would quickly rise from unity to $\Delta_b=\rho_b/\rho_c\gg 1$,
where $\rho_c$ and $\rho_b$ are respectively the average cosmological
energy density and the density of PBHs in the cluster (bunch). It looks
reasonable that  the density contrast of the evolved cluster could rise up
to ${\Delta_b = 10^{5}-10^{6}}$, as in the contemporary galaxies.
After the size of the cluster stabilized, 
the number density of PBH, $n_{BH}$, as well as 
their mass density, $\rho_{BH}$,
would be constant too. But the density contrast, $\Delta_b$ 
would continue to rise as $(t/t_1)^2$ because $\rho_c$ drops down as $1/t^2$. 
From time $t=t_1$ to $t=\tau_{BH}$ the density contrast would
additionally rise by the factor:
\be
\Delta (\tau_{BH})= \Delta_b \left(\frac{\tau_{BH}}{t_1}\right)^2\,.
\label{Delta-of-tau}
\ee
This rise is associated with the drop of the average cosmological energy density,
$\rho \sim 1/t^2$, but not with the absolute rise of $\delta\rho$.
This effect is absent in the present day universe because the time when $\Delta $ reached 
unity was close to the present universe age. 

GWs could be generated in the processes of PBH scattering in the high density clusters and,
in particular, the GW emission could proceed from the PBH binaries. 
Both processes are strongly  enhanced in the clusters.
The probability of scattering and binary formation rate
are proportional to the square of the number density of PBHs, $n_{BH}$. However, the net 
effect on the cosmological energy density of the emitted GWs is linear in $n_{BH}$ because
it is normalized to the total cosmological energy density.

The cross-section of the graviton bremsstrahlung was calculated in 
ref.~\cite{ref:brems} for the case of two spineless particles 
(here black holes) with masses $m$ and  $M$ under assumption that $m\ll M$. 
In non-relativistic approximation, the differential cross section is:
\begin{equation}
\mathrm{d}\sigma=\frac{64M^2m^2}{15m_{pl}^6}\frac{\mathrm{d}\xi}{\xi}
\left[ 5\sqrt{1-\xi} +\frac{3}{2} (2-\xi) \ln
  \frac{1+\sqrt{1-\xi}}{1-\sqrt{1-\xi}}\, \right]\,,
\label{sigma-brems}
\end{equation}
where $\xi$ is the ratio of the emitted graviton frequency,
$\omega=2\pi f$, to the kinetic energy of the incident black hole,
i.e. $\xi = 2m \omega / {\bf p}^2$.
In what follows we will use this expression for a simple estimate assuming that it
is approximately valid  for $m\sim M$. 

The energy density of gravitational waves emitted at the time interval $t$ and $t+\mathrm{d}t$ 
in the  frequency range $\omega$ and $\omega+\mathrm{d}\omega$ is given by
\be
\frac{\mathrm{d}\rho_{GW}}{\mathrm{d}\omega} = 
 v_{rel} n_{BH}^2 \omega\left(\frac{\mathrm{d}\sigma}{\mathrm{d}\omega}\right)\mathrm{d}t\, ,
\label{d-dot-rho-domega}
\ee
where $n_{BH}$ is the number density of PBH and $v_{rel}$ is their
relative velocity. The latter is close to the virial velocity of PBHs in the cluster and can be about
0.1. As noted by the authors of ref.~\cite{ref:brems},  Weizs{\" a}cker-Williams approximation
is not valid. This means that there could be some difference between classical and
quantum graviton emission.

The graviton bremsstrahlung proceeded till the PBH evaporation. Hence to find the total energy of
the produced gravitons we need to integrate their energy spectrum over frequencies and redshift
from $\tau_{BH}$ down to the moment of the cluster formation. Thus we obtain for the
cosmological energy fraction of GWs:
\be
{\Omega_{GW}^{(brems)} (\omega_{max}, \tau_{BH})}
\,\approx 16 Q\,  \left(\frac{v_{rel}}{0.1}\right)\,
\left(\frac{\Delta}{10^5}\right)\,
\left(\frac{N_{eff}}{100}\right)\, \left(\frac{\omega_{max}}{M}\right)\,.
\label{rho-brems}
\ee
Here coefficient $Q>1$ reflects the uncertainty in the 
cross-section due to the unaccounted for Sommerfeld enhancement~\cite{ref:sommer}.
Note that $\Delta$ may be considerably larger than $10^5$. For the details see 
ref.~\cite{ref:de}.

The frequency $f_*$ of GW produced at time $t_*$ during PBH evaporation, is redshifted down to 
the present day value, $f$, as:
\be 
f = f_*\left[\frac{a(t_*)}{a_0}\right]=0.34\, f_*\, \frac{T_0}{T_{*}} \left[\frac{100}{g_S (T_*)}\right]^{1/3}\,,
\label{Omega-brems-today}
\ee
where $T_0= 2.725$ K is the
temperature of the cosmic microwave background 
radiation at the present time,  $T_*\equiv T(t_{*})$ is the plasma
temperature at the moment of radiation of the gravitational waves, and
$g_S(T_*)$ is the number of species contributing to the entropy of the primeval plasma at temperature $T_*$.
It is convenient to express $T_0$ in frequency units, $T_0 = 2.7 \,{\rm  K} = 5.4\cdot 10^{10}$ Hz.

The density parameter of the gravitational waves at the present time 
is related to cosmological time $t_*$ as: 
\begin{equation}
\Omega_{GW}(t_0)=\Omega_{GW}(t_{*})\left(\frac{a(t_{*})}{a(t_0)}\right)^4\left(\frac{H_{*}}{H_0}\right)^2\,,
\label{Omega-GW-of-H} 
\end{equation}
where $H_0 =100 h_0 $~km/s/Mpc is the Hubble parameter and $h_0=0.74\pm 0.04$ \cite{ref:data-h}.

Using expression for redshift (\ref{Omega-brems-today}) and taking the emission time $t_*=\tau_{BH}$ we obtain:
\begin{equation}
\Omega_{GW}(t_0)=1.67\times 10^{-5} h_0^{-2}\left(\frac{100}{g_S(T_{BH})}\right)^{1/3}\Omega_{GW}(\tau_{BH})\,.
\label{generalexpression}
\end{equation}
Now we find that the total density parameter of gravitational waves 
integrated up to the maximum frequency is:
\be
h_0^{2}\Omega_{GW} (t_0) \approx 0.6\cdot 10^{-21}\,K
\left(\frac{10^5\,\textrm{g}}{M}\right)^{2}\,,
\label{Omega-brems-0}
\ee
where $K$ is the numerical coefficient:
\be
K = \left(\frac{v_{rel}}{0.1}\right)\,
\left(\frac{\Delta}{10^5}\right)\,
\left(\frac{N_{eff}}{100}\right)\,
\left(\frac{Q} {100}\right)\,
\left(\frac{100}{g_S (T_{BH})}\right)^{1/3}\,.
\label{kappa}
\ee
Presumably $K$ is of order unity but since 
$\Delta$ may be much larger than $10^5$, see eq. (\ref{Delta-of-tau}), $K$ may also be large.

Classical emission of GW at the scattering of non-relativistic bodies
is well described in quadrupole approximation.  If the minimal
distance between the bodies is larger than their gravitational radii,
the energy of gravitational waves emitted in a single scattering process is equal to:
\begin{equation}
\delta E_{GW}=\frac{37\pi}{15}\frac{M^2m^2v}{b^3m_{Pl}^6},\qquad v\ll 1\,,
\label{delta-E-nonrel}
\end{equation}
where $b$ is the impact parameter.

The differential cross-section of the gravitational scattering of two PBHs in non-relativistic regime, 
$q^2\ll 2M^2$, is:
\be
\mathrm d\sigma = \frac{M^2}{m_{Pl}^2}\, \frac{\mathrm dq^2}{q^4}
=\frac{2M^2}{m_{Pl}^2} b \mathrm db\,  
\label{d-sigma-dq2}
\ee
and the rate of the energy emission by the GWs is given by:
\be 
\mathrm{d}\rho_{GW} = \frac{74 \pi v_{rel}} {15} \,\rho_{BH}^2\,
\frac{M^4}{m_{Pl}^8} \,\frac{\mathrm d\omega}{2\pi}\, \mathrm dt\,.
\label{d-dot-rho-GW}
\ee
The energy density parameter of GW at the moment of BH evaporation can be obtained integrating 
this expression over time and frequency. Thus we obtain:
 \be
\Omega_{GW} (\tau_{BH})= 2\cdot 10^{-10}\left(\frac{v_{rel}}{0.1}\right)^2 
\left(\frac{\Delta_b}{10^5}\right)\,
\left(\frac{N_{eff}}{100}\right)\,
\left(\frac{10^5\,\textrm{g}}{M}\right)\,.
\label{Omega-GW-of-t}
\ee  
If we allow for $b \sim r_g$, the energy
density of GWs at the moment of PBHs evaporation might be comparable to unity.

Now we can calculate the relative energy density of GWs per logarithmic frequency interval 
at the present time:
\begin{equation}
\Omega_{GW} (f; t_0)
\equiv \frac{1}{\rho_c}\,\frac{\mathrm{d}\rho_{GW}}{\mathrm{d}\ln f} \approx
2.4\cdot 10^{-12}\alpha '\, \left(\frac{f}{\textrm{GHz}}\right)\,\left(\frac{10^5\,\textrm{g}}{M}\right)^{1/2}\,,
\label{Omega-GW-of-f-t_0}
\end{equation}
where $\alpha '$ is the coefficient at least of order of unity:
\begin{equation}
\alpha '= \left(\frac{v_{rel}}{0.1}\right)
\left(\frac{\Delta_b}{10^5}\right) 
\left(\frac{N_{eff}}{100}\right)^{3/2}
\left(\frac{100}{g_S(T_{BH})}\right)^{1/4}\,.
\end{equation}
It may be much larger, if $\Delta_b\gg 10^5$.

More efficient mechanism of GW emission may be radiation from the PBH
binaries, if their number in the high density clusters is sufficiently high. To form
the binary bound state PBHs should sufficiently cool down losing their kinetic
energy.  The cooling could be achieved by the energy loss to the gravitational wave 
radiation discussed above and by the dynamical friction~\cite{ref:dyn-fric}. 
A particle moving in the cloud of other particles would
transfer its energy to these particles due to their gravitational interaction. 
However, one should keep in mind that the case of
dynamical friction is essentially different from the energy loss due to
gravitational radiation. In the latter case the energy leaks out of the
system cooling it down, while dynamical friction does not change the
total energy of the cluster. Nevertheless a particular pair of black holes moving 
toward each other with acceleration may transmit their energy to the
rest of the system and became gravitationally captured forming a binary.

The dynamical friction time was estimated in ref.~\cite{ref:bambi-fric}. In both
cases $v>\sigma$ and $v <\sigma$, where $\sigma$ is the velocity dispersion,
the characteristic time was of the order of:
\be
\tau_{DF}  
\approx \left(\frac{\sigma}{0.1}\right)^3 \left[\frac{25}{\ln (10^{-6}/\Omega_p)}\right]
\left(\frac{100}{N_{eff}}\right) \left(\frac{M}{1\,{\rm g}}\right)
\left(\frac{10^6}{\Delta}\right)\tau_{BH}\,.
\label{tau-DF}
\ee
For PBH masses below a few grams  dynamical friction would be an efficient mechanism 
of PBH cooling leading to frequent binary formation. Moreover,
dynamical friction could result in the collapse of small PBHs into much
heavier black hole. Even the whole high density cluster of PBHs could form a single black
hole. These processes of heavier black hole formation
would be accompanied by a strong burst of gravitational radiation.

The emission of  GWs from a binary results in the energy loss which is compensated
by a decrease of the radius of the binary and of the rotation period. 
As a result the system goes into the so called inspiral regime. 
Ultimately the two rotating bodies coalesce and produce a burst of the 
gravitational waves. To reach this stage 
the characteristic time of the coalescence should be shorter than the life-time of the system. 
In our case it is the life-time of PBH with respect to the evaporation.
The coalescence time of the binary made of two BH with masses $M_1$ and $M_2$
can be easily calculated, see e.g. book~\cite{ref:ll2}:  
\be
\tau_{co} = \frac{5 R_0^4\,m_{Pl}^6}{256 M_1 M_2 (M_1+M_2)}\,,
\label{t-coaelescence}
\ee
where $R_0$ is the initial radius of the binary. This result is true for a circular orbit of the
binary. In the case of elliptic orbit the eccentricity drops down due to GW emission
and the system approaches to the circular one. We may use eq. (\ref{t-coaelescence})
for an order of estimate of the life-time of the binary.

There are two interesting limiting cases, when $\tau_{co} \gg \tau_{BH}$ and vice versa.
In the first case the stationary orbit approximation is valid and each binary emits GWs with fixed
frequency equal to twice the orbital frequency and the frequency spectrum is determined by
the distribution of the binaries on their radius. As is shown in ref.~\cite{ref:de}, if the stationary 
regime was realized, the spectral density parameter today would be:
\be
\Omega_{GW}^{(stat)} (f; t_0) \approx 10^{-8} \epsilon 
\left[\frac{N_{eff}}{100}\right]^{2/3}\left[\frac{100}{g_S(T(\tau_{BH}))}\right]^{1/18} 
\left[\frac{M}{10^5\,{\rm g}}\right]^{1/3}\,\left[\frac{f}{{\rm {GHz}}}\right]^{10/3}\,,
\label{Omega-upper-limit}
\ee
where $\epsilon$ is the fraction of binaries with respect to the total number of PBHs
in the cluster and $g_S (T(\tau_{BH})$ is the number of the entropy degrees of freedom
at the moment of PBH evaporation when the plasma temperature was equal to $T(\tau_{BH})$.
Here the possibly weak redshift dilution of GWs by the factor $(\tau_{co}/\tau_{BH})^{2/9}$
is neglected.
   
If the system goes to the inspiral phase, then we would expect today  
a continuous spectrum in the range from $f_{min}\sim
 10^7 \textrm{Hz} $ to $f_{max}\sim 3\cdot 10^{14}\,\textrm{Hz}$. 
However if we take into account the redshift of the early formed binaries from the moment of their 
formation to the PBH decay, the lower value of the frequency may move to about 1 Hz.

PBHs could also directly produce gravitons by evaporation.
The total energy emitted by BH per unit time and frequency $\omega$ (energy) of the emitted particles, 
is approximately given by the equation (see, e. g. book \cite{ref:frol-nov}):
\begin{equation}
\left(\frac{\mathrm{d}E}{\mathrm{d}t\mathrm{d}\omega}\right)=
\frac{2N_{eff}}{\pi}\,\frac{M^2 }{m_{Pl}^4}\frac{\omega^3}{e^{\omega/T_{BH}}-1}\,,
\end{equation}
where $T$ is the BH temperature (\ref{T-BH}). 
Due to the impact of the gravitational field of BH on the propagation of the evaporated particles, their spectrum 
is distorted \cite{ref:page} by the so called grey factor $g(\omega)$, but we disregard it in what follows. 

The frequency spectrum of the evaporated gravitons is not thermal because of the different redshifts
in the course of the evaporation. According to the calculations of ref.~\cite{ref:de}
the spectral density parameter of GWs at $t=\tau_{BH}$ is equal to:
\be
\Omega_{GW}(\omega_*; \tau_{BH})\approx
\frac{ 2.9\cdot 10^3 M^4 \omega_*^4}{\pi\, m_{Pl}^8}\,I\left(\frac{\omega_*}{T_{BH}}\right)\,,
\label{evaporation-BH}
\ee
where 
\begin{equation}
I\left(\frac{\omega_*}{T_{BH}}\right) = \int_{0}^{z_{max}} \frac{ \mathrm dz \left(1+z \right)^{1/2}}
{\exp\left[(z+1) \omega_*/T_{BH} \right] -1}\,,
\end{equation}
and
\begin{equation}
1+z_{max} = \left(\frac{\tau_{BH}}{t_{eq}}\right)^{2/3}\left(\frac{t_{eq}}{t_p}\right)^{1/2}=\left(\frac{32170}{N_{eff}}\right)^{2/3}\left(\frac{M}{m_{Pl}}\right)^{4/3}\Omega_p^{1/3}\,,
\end{equation}
With respect to the thermal spectrum, spectrum (\ref{evaporation-BH}) 
has more power at small frequencies due to 
redshift of higher frequencies into lower band and less power at high $\omega_*$. 

The spectral density of the evaporated gravitons today would be: 
\begin{equation}
\Omega_{GW}(f; t_0)=2.7\cdot 10^{-27}\left(\frac{N_{eff}}{100}\right)^2\left(\frac{10^5\,\textrm{g}}{M}\right)^2\left(\frac{f}{10^{10}\,\textrm{Hz}}\right)^4\cdot I\left(\frac{2\pi\cdot f}{T_0}\right)\,,
\end{equation}
where $T_0$ is the BH temperature redshifted to the present time:
\begin{equation}
T_0= 4.5\cdot 10^{15}\,\textrm{Hz}\left(\frac{100}{g_S(T_{BH})}\right)^{1/12}
\left(\frac{100}{N_{eff}}\right)^{1/2}\left(\frac{M}{10^5\,\textrm{g}}\right)^{1/2}\,.
\end{equation}

The mechanisms of GWs generation considered here could 
create quite high cosmological fraction of the energy density of the relic gravitational waves 
at very high frequencies. Unfortunately at the lower part of the spectrum $\Omega_{GW}$ 
significantly drops down making such GWs outside the reach of LISA or LIGO.
Still the planned interferometers DECIGO/BBO and detectors based on the resonance
graviton-photon transformation could be sensitive to the predicted high
frequency GWs.

\end{document}